\documentclass[%
 reprint,
 amsmath,amssymb,
 aps,
]{revtex4-1}

\usepackage{graphicx}
\usepackage{dcolumn}
\usepackage{bm}

\usepackage{natbib}

\begin{document}

\preprint{APS/123-QED}

\title{Standing fast: Translation among durable representations using evanescent representations in upper-division problem solving}

\author{Nandana Weliweriya}
\author{Tra Huynh}
\author{Eleanor Sayre}%
\affiliation{%
 Department of Physics, College of Arts and Sciences, Kansas State University\\
}%




\date{\today}

\begin{abstract}
Mastering problem solving requires students to not only understand and apply physics concepts but also employ mathematics and mathematical representations (sketches, diagrams, graphs, gestures, equations and spoken language) skillfully. As part of a larger project to investigate problem solving processes among upper division physics students, we investigate how students coordinate among multiple representations while solving problems. Data for this study is drawn from an upper-division Electromagnetism I course, where students engage in individual oral exams. We do moment-by-moment analysis of students' problem solving to see how they translate between durable representations (diagrams, written mathematical equations) with the help of evanescent representations (gestures, words); and how they build up durable representations where they can ``stand fast" later. In this paper, we present the case of Larry as an exemplary case for problem solving. Larry starts from a durable representation (diagram) and builds up from there using evanescent representations (gestures and words), standing fast on the diagram. He later translates to a different kind of durable representation (mathematics), where he reasons and answers the original problem. 
\end{abstract}

\pacs{Valid PACS appear here}
\maketitle


\section{\label{sec:level1}Introduction}

In the course of problem solving, students use a series of representations to make sense of the problem, build a solution, and communicate it. In this paper, we are concerned with how students build up and use representations in problem solving, focusing on a single case at the upper-division level.  

At their most basic, representations are either external or internal objects that stand for something else.  A descriptive representation consists of symbols describing an object.  Spoken or written texts and mathematical equations are examples of descriptive representations.  On the other hand, a depictive representation consists of iconic signs. Pictures or physical models are examples of depictive representations.  Most of the prior research in student understanding and use of multiple representations (e.g. \cite{rosengrant_case_2006,flood_paying_2015,gire_graphical_2012})   takes up this categorization scheme, either implicitly or explicitly. Using  multiple representations plays a critical role in the effectiveness of the interactive engagement between students and instructors in learning environments\cite{rosengrant_case_2006, fredlund_exploring_2012}. Gestures can be a productive medium for representation, visualization and interaction in chemistry problem solving, particularly in how students use gestures to describe molecular geometry in their own style to develop a technical language\cite{flood_paying_2015}. In upper-division electricity and magnetism classes, students find vector field manipulation is easier with algebraic representational approach\cite{gire_graphical_2012}, though students need to use graphical representations to interpret the difference between components and coordinates. In addition, translating between different representations may enhance students' sense-making abilities\cite{gire_graphical_2012} .

Alternately, we can categorize representations as to whether they are durable or evanescent.  A persistent (durable) representation ``leaves a trace of its production in the medium in which it was produced"\cite{fredlund_exploring_2012}. Durable representations stay in place and can be easily returned to, such as diagrams, graphs, written text, and written mathematical equations. In contrast, evanescent representations are temporary and disappear when not in use, such as verbal and facial expressions, as well as physical gestures.


We take up the language of ``standing fast'' to describe how students use durable and evanescent representations in their problem solving.  The term ``stand fast" refers to remaining firmly in the same position or keeping the same opinion. In the context of problem solving, people stand fast  on ``something that a person or group of people can call upon and use without hesitation or further questioning"\cite{fredlund_exploring_2012}. This is when in a conversation all the participants agree on a presented idea or an explanation \cite{wickman_learning_2002}. We relate this idea of standing fast to the process of students' reasoning while solving problems. As students progress and develop their solution and the reasoning behind it, there is a point that students no longer hesitate and the instructor or other peers can accept his/her solution or the explanation without any further discussion. We see students switch back and forth between explanations to stand fast to a representation until they find it is insufficient for a complete answer. 

Fredlund \cite{fredlund_exploring_2012} claims that durable representations can help students to stand fast, as students keep returning to them for sense-making while they build evanescent representations on top of these durable representations. 

\begin{figure*}[t]
\centering
\setlength\fboxsep{0pt}
\setlength\fboxrule{0pt}
\fbox{\includegraphics[width=0.7\textwidth]{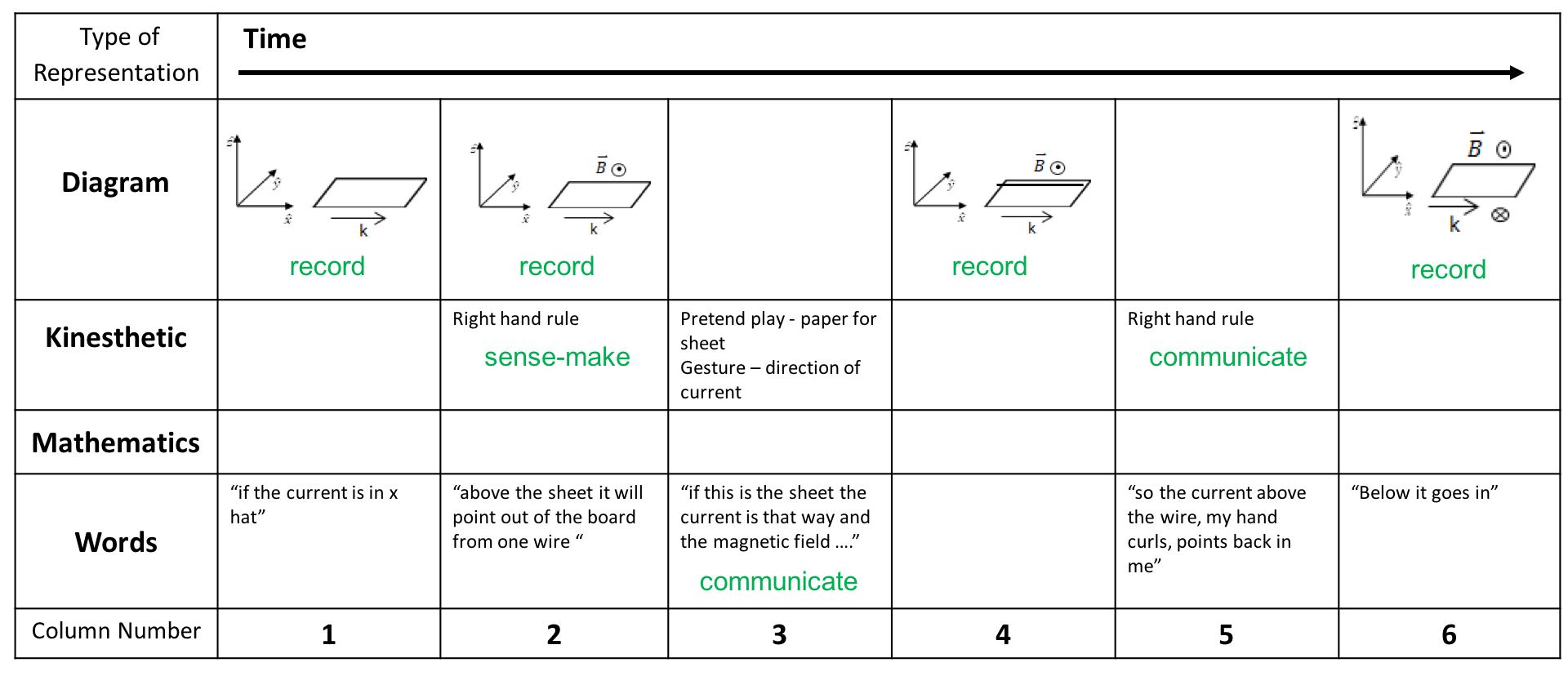}}
\caption{\label{figure1}Larry works to figure out the magnetic field direction}
\label{f7}
\end{figure*}
%
%

In this paper, we explore two research questions: 
how can problem solving be described as a process of transitioning between durable representations to make new meanings? How do students build meanings onto durable representations using the evanescent representations? To explore these questions, we present the case study of a student solving a problem during an oral exam in an upper-division Electromagnetism I course.

\section{\label{sec:3evel1}Method}
Data for this study is from an undergraduate upper- division Electromagnetism I course usually taken by seniors with an enrollment about 20 students. The class was taught by a white female instructor who had taught the same course at a different small institution. This particular course covers the first seven chapters of Introduction to Electrodynamics (4th Edition) by David J. Griffiths and during the class students work on tutorials and small group problem solving. In order to answer our research questions, we do a moment-by-moment analysis of students' videos to see how students build up and translate between representations and how they stand fast. In this paper, we present the case of Larry, a white male student, who engages in problem solving for an individual oral exam during the later part of the course. Larry is a strong student, near the top of his class.  We selected him because he is unusually verbal in oral exams compared to his peers and he makes a good example of how students coordinate multiple representations.

\section{\label{sec:4evel1}Larry's representations}

Larry is working on a problem for his oral exam in his Electromagnetism I course. During this episode, Larry starts by developing a diagram (durable representation) and spends time building meaning through gestures and words (evanescent representations). After recording the developed meaning on the diagram, Larry moves to mathematics (durable representation) in the later parts of his solution to calculate the magnitude. The episode starts with instructor posting the problem, \textit{Suppose you had an infinite sheet which carries current  $k$ equal to some constant ($k = \alpha \hat{x}$ ). 
What's that look like? What kind of a physical scenario is that?} 

To solve this problem, one can use the right hand rule to find the direction of the magnetic field created by the sheet and to find the magnitude we could use the Ampere's law ($\oint B\cdot\mathrm{d}l =\mu_0 I_{enc}$). 

However, Larry starts by drawing the sheet of current along with a Cartesian coordinate system to represent the given current with direction, noting that ``so if the current is in x hat ($ \hat{x}$)\dots"  Larry picks the mathematical information (direction of the current: $ \hat{x}$  ) and uses it to develop a diagram (durable representation) to start with (\textbf{Figure 1: column 1}). After recording the sheet of current, Larry employs the right hand rule (\textbf{Figure 1: column 2}) with the current on the diagram to come to a conclusion and records the direction of the magnetic field direction on the board (\textbf{Figure 1: column 2}).

\begin{description}
\item[Larry] above the sheet it would be pointing out of the board. So I think it would be true for the rest of the sheet as well. 
\end{description}

Larry is not provided with a diagram, so he has to come up his own to progress. He uses evanescent representation gestures to improve his diagram and to build new meaning and then records the developed meaning on his diagram. At this point, we see Larry uses the developed diagram to reason with no hesitation or further questions. He uses the improved durable representation as his first place to stand fast.

\begin{figure*}[t]
\centering
\setlength\fboxsep{0pt}
\setlength\fboxrule{0pt}
\fbox{\includegraphics[width=0.6\textwidth]{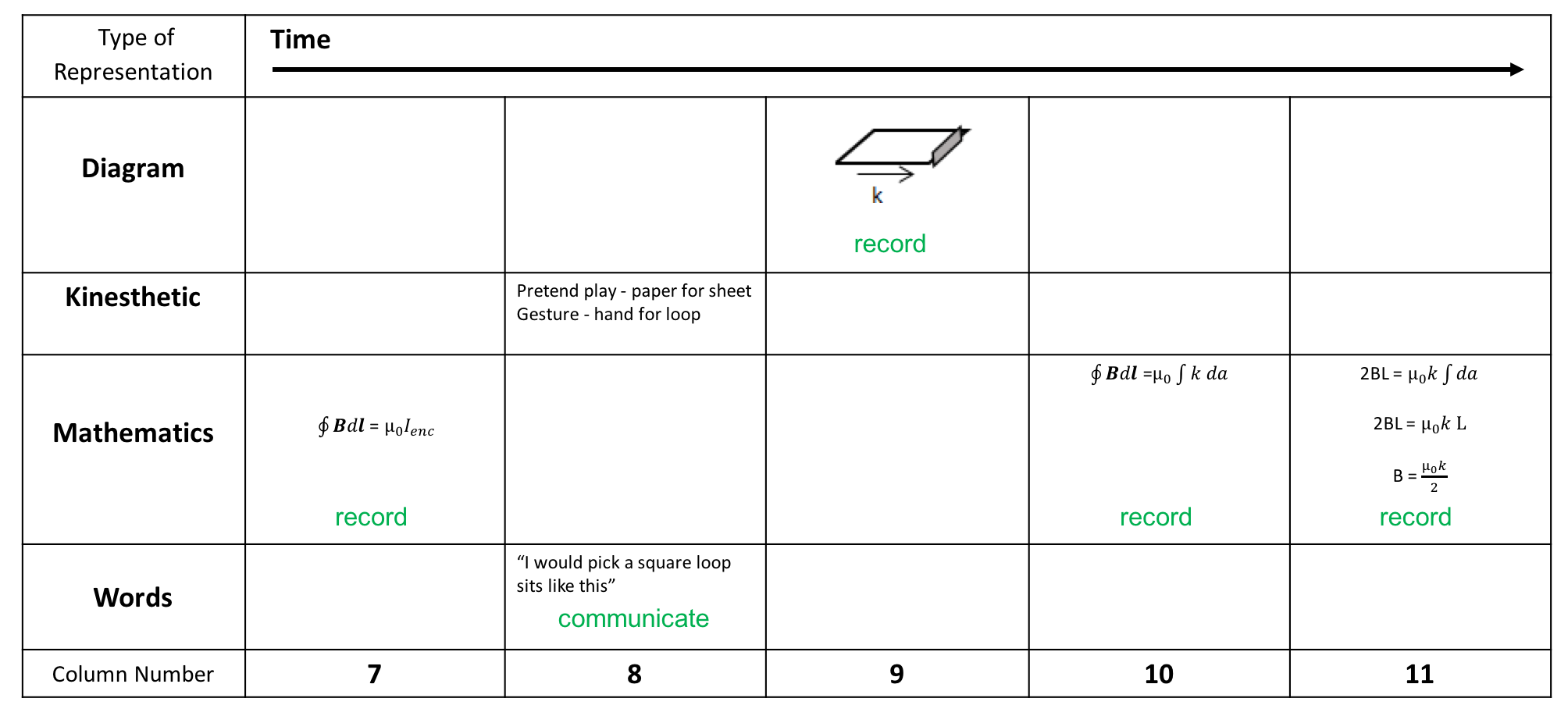}}
\caption{\label{figure2} Larry works to figure out the magnitude of magnetic field}
\label{f7}
\end{figure*}

Up to this point Larry has worked on the problem using the diagram on board, but for further clarification instructor brings in a different representation: a sheet of paper.  The physical paper is durable; the hand gestures showing the direction of the current and verbal accompaniment are evanescent. Throughout the solution we see Larry keep referring to the same sheet of paper as a sheet of current, so we interpret the representation as a combination of the physical paper and the meanings associated with it as a sheet of current. This whole representation persists throughout the problem.  According to our definition mentioned in section one, within the context of this problem we assign the sheet of paper to be a durable representation.

\begin{description}
\item[Instructor] Ok, So if this is the sheet, if the current is this way (uses the hand to show the direction)
%
%
%
%
%
\item[Larry] Okay, so if it is the sheet, the current is going this way and looking at a point above it, then from one wire, magnetic field will point in that way.
\end{description}

Larry starts using the new durable representation, the sheet of paper, to figure out the direction of the magnetic field created by a single wire (\textbf{Figure 1: column 3}). Where Larry builds new meaning on the top of durable representation, using  evanescent representations, gestures and words.
%
%

After determining the magnetic field created by a single wire, Larry moves to consider a different reference point for a current carrying wire. To do so he goes back to his own on-board problem (his original durable representation) and draws (\textbf{Figure 1: column 4}) a line right in the middle of the sheet (to represent a current carrying wire). 

Larry then stretches his arm on the board to show a point in the $y$ direction, then applies the right hand rule again (\textbf{Figure 1: column 5}). Here, Larry goes back to his original durable representation and improves his diagram. Then he builds the evanescent representation, gestures on top of it to determine the magnetic field created by a different current wire. 

The instructor introduces the idea of a mirror current,
\begin{description}
\item[Instructor] What’s about the one, that mirror, so that wire way far back, the mirror wire way far forwards?
\end{description}
Larry replies quickly that ``[his] finger is pointing up'', concluding that the direction of magnetic field is ``out" above and ``in" below and records that on the board (\textbf{Figure 2: column 6}).

Larry does lots of sense-making using the evanescent representations to develop his original durable representation. Though the sheet of paper is durable, Larry does not stand fast on it; rather he uses it to improve upon and make more sense of his diagram. 
He uses the developed diagram to stand fast for the second time and answer the original question about the direction of the magnetic field by the sheet of current. 

After figuring out the direction of magnetic field, the instructor asks Larry to calculate the magnitude. Because Larry could not remember the equation for the magnetic field from a single wire, instructor suggests using Ampere's law. Larry right away starts by recording the formula (\textbf{Figure 2: column 7}). So far during the first section of his solution, Larry built and used his diagram for reasoning, but just the diagram is insufficient to calculate the magnitude. Larry translates his prior work to mathematics, which is a new durable representation. 

The instructor gives him a hint to pick an Amperian loop, and Larry uses his hand to describe the loop (\textbf{Figure 2: column 8}), but he switches back to the sheet of paper to do so,

\begin{description}
\item[Larry] uh\dots yes. And I would pick a loop. Uh\dots I guess I would pick a square loop sits like this (again uses the sheet of paper and use fingers to show a cut at the middle of the sheet).
%
%

\end{description}

Larry gestures to show a loop which is parallel to the edge of paper. Here we see Larry is using evanescent representations along with the sheet of paper to progress and build up mathematics.   

The instructor suggests that he draw the Amperian loop on the board, asking ``So how big is your loop? Draw your loop.''  Even though Larry gestures to make the size and orientation of his chosen loop successfully (Figure 3), he is not sure how to draw it on his diagram.  

To figure out how to draw it, Larry uses gestures and even comes back to the sheet of paper to help him with drawing. After a while Larry records the loop on the diagram (\textbf{Figure 2: column 9}) and labels the sides of the loop with $l$ and $w$. Here we see Larry is referring back to the sheet of paper and improves his diagram (durable representation) using gestures and words (evanescent representations) to finish the mathematical calculation for the magnitude of the magnetic field.

\begin{figure}[t]
\centering
\includegraphics[scale=.42]{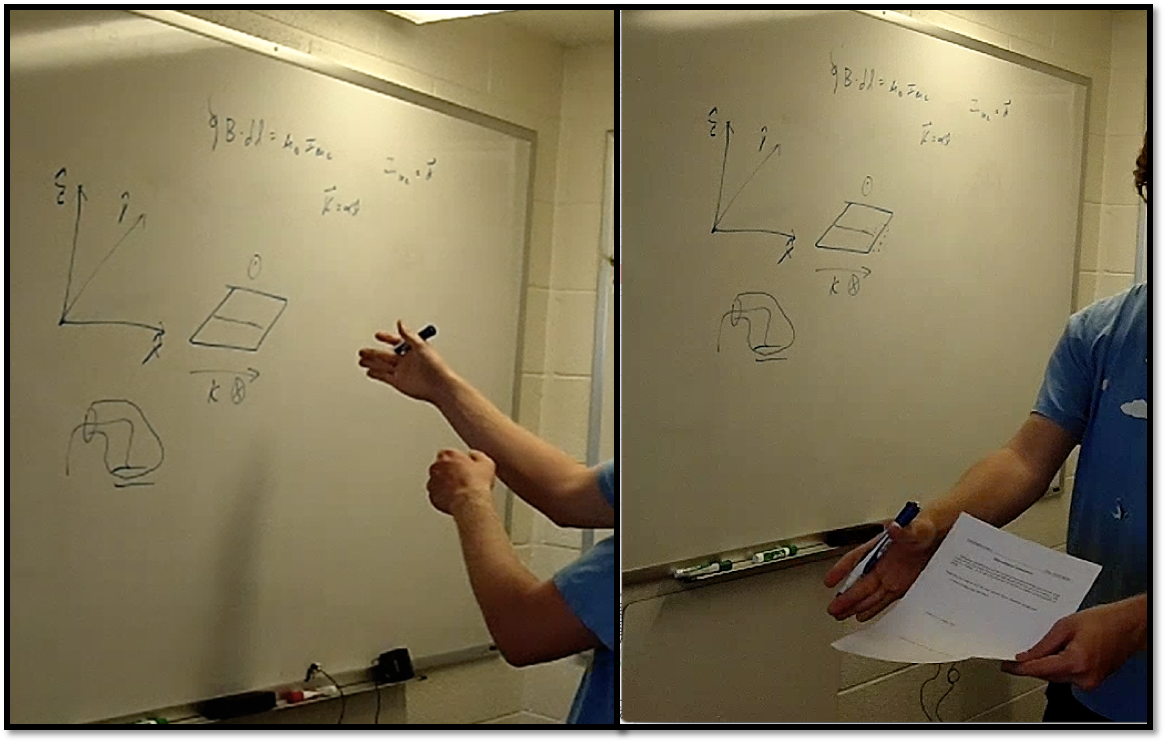}
\caption{\label{figure3} Larry gestures on the board and uses the sheet of paper to show how he picks the Amperian loop}
\end{figure}

Once Larry draws the loop, the instructor asks a question about the current piercing the loop. Larry answers those questions by developing the mathematics on the board (\textbf{Figure 2: column 10}) and reasons that for the right hand side of the integral ($\oint B\cdot\mathrm{d}l =\mu_0 \int k\cdot\mathrm{d}a$). 

Communication with the instructor leads Larry to picks a direction for his loop and records it as counter-clockwise. After this step, Larry takes a little time to compute the right side. With a little help from instructor Larry reaches the final answer (\textbf{Figure 2: column 11}). Here Larry stands fast on the mathematics to answer about the magnitude of the magnetic field. 

During the second half of this problem, Larry translates from diagram to mathematics. He uses the durable representation, sheet of paper and evanescent representations, gestures and words to amend mathematics.  After developing the mathematics representation, he uses it to stand fast in the later part of his solution.



\section{\label{sec:5evel1}Discussion and Conclusion}

The case of Larry shows a good example of how students translate between durable representations, with the help of evanescent representations. In summary, we see Larry starts using one kind of durable representation (diagram), then translates to a different kind of durable representation (mathematics) using the evanescent representations (gestures and words) to reason better and answer the original problem. Larry stands fast on the first durable representation (diagram), which he builds in the beginning. Later in his solution, Larry stands fast on mathematics.

This particular problem required the student to start from the diagram and move to the mathematics, so is a great case to investigate how a student may transition between two durable standing fast representations.  There exist other, simpler problems which may only require one representation to stand fast upon; conversely, there are more complicated, research-like problems which may require a long series of representations on which to stand fast.  

We have evidence for Larry building new meaning on durable representations using evanescent representations. Larry brings in evanescent representations to amend his first durable representation: diagram, because it was not sufficient for further reasoning. Larry spends most of his time building his durable representation: the diagram, where he stands fast before translating to other durable representation: mathematics, when calculating the magnitude of the magnetic field. Larry uses the sheet of paper along with gestures and words (evanescent representations) to build up mathematics and also to bridge the two durable representations.

Translating between different classes of representations might improve student understanding and how the combinations of representations could be effective for visualization and interaction in other disciplines\cite{flood_paying_2015}.  As we think about classifying representations, our choice of categorization scheme impacts what translating between representations entails.  Building meaning on durable representations using evanescent ones can be fruitful for problem solving; choosing it as a categorization scheme focuses our efforts on the flow of representations in problem solving and how students stand fast. 

\section{\label{sec:6evel1}Acknowledgements}
This work is supported by National Science Foundation grant 1430967.  We are deeply grateful to the members of the Mathematization group at KSUPER, our copyeditors, and three anonymous reviewers whose comments improved this work.

\bibliographystyle{ieeetr.bst}
\bibliography{My_Library}

\end{document}